\documentclass[11pt]{article}

\usepackage[a4paper,margin=1in]{geometry}
\usepackage[T1]{fontenc}
\usepackage[utf8]{inputenc}
\usepackage{lmodern}
\usepackage{microtype}
\usepackage{graphicx}
\usepackage{float}
\usepackage{booktabs}
\usepackage{tabularx}
\usepackage{array}
\usepackage{enumitem}
\usepackage{xcolor}
\usepackage[numbers,sort&compress]{natbib}
\usepackage[colorlinks=true,linkcolor=blue!50!black,citecolor=blue!50!black,urlcolor=blue!60!black]{hyperref}
\usepackage{xurl}
\usepackage[most]{tcolorbox}

\newcolumntype{Y}{>{\raggedright\arraybackslash}X}
\newcolumntype{P}[1]{>{\raggedright\arraybackslash}p{#1}}

\setlist[itemize]{leftmargin=1.4em}
\newtcolorbox{namedbox}[1]{
  enhanced,
  breakable,
  colback=black!3,
  colframe=black!55,
  boxrule=0.45pt,
  arc=2pt,
  left=6pt,
  right=6pt,
  top=6pt,
  bottom=6pt,
  title=\textbf{#1}
}

\newcommand{\figureplaceholder}[2]{%
  \IfFileExists{#1}{%
    \includegraphics[width=\linewidth]{#1}%
  }{%
    \fbox{%
      \parbox[c][#2][c]{0.93\linewidth}{%
        \centering
        Placeholder for figure file\\[0.35em]
        \texttt{#1}
      }%
    }%
  }%
}

\title{The Semi-Executable Stack: Agentic Software Engineering and the Expanding Scope of SE\thanks{This paper is the write-up of Robert Feldt's keynote ``Agentic Software Engineering Will Eat the World: AI-Based Systems as the New Operating System of Society'' given at the Agentic Engineering 2026 workshop, Rio de Janeiro, Brazil, April 14, 2026. The co-authors were especially important in discussions leading up to the keynote and were invited to comment and contribute to the write-up.}}
\author{Robert Feldt$^{1}$, Per Lenberg$^{1}$, Julian Frattini$^{1}$, Dhasarathy Parthasarathy$^{2}$\\$^{1}$Chalmers University of Technology, Gothenburg, Sweden\\$^{2}$Volvo Group, Gothenburg, Sweden\\Correspondence: \texttt{robert.feldt@chalmers.se}}
\date{April 2026}

\begin{document}
\maketitle

\begin{abstract}
AI-based systems, currently driven largely by LLMs and tool-using agentic harnesses, are increasingly discussed as a possible threat to software engineering. Foundation models get stronger, agents can plan and act across multiple steps, and tasks such as scaffolding, routine test generation, straightforward bug fixing, and small integration work look more exposed than they did only a few years ago. The result is visible unease not only among students and junior developers, but also among experienced practitioners who worry that hard-won expertise may lose value. This paper argues for a different reading. The important shift is not that software engineering loses relevance. It is that the thing being engineered expands beyond executable code to \emph{semi-executable artifacts}---combinations of natural language, tools, workflows, control mechanisms, and organizational routines whose enactment depends on human or probabilistic interpretation rather than deterministic execution. \emph{The Semi-Executable Stack} is introduced as a six-ring diagnostic reference model for reasoning about that expansion, spanning executable artifacts, instructional artifacts, orchestrated execution, controls, operating logic, and societal and institutional fit. The model helps locate where a contribution, bottleneck, or organizational transition primarily sits, and which adjacent rings it depends on. The paper develops the argument through three worked cases, reframes familiar objections as engineering targets rather than reasons to dismiss the transition, and closes with a preserve-versus-purify heuristic for deciding which legacy software engineering processes, controls, and coordination routines should be kept and which should be simplified or redesigned. This paper is a conceptual keynote companion: diagnostic and agenda-setting rather than empirical.
\end{abstract}

\section{What Agentic Software Engineering Changes}

This paper uses \emph{agentic software engineering} to mean software engineering whose practice and artifacts increasingly involve autonomous or semi-autonomous AI systems that act through tools, maintain state across steps, and coordinate with humans and other agents. In current practice, these systems are usually built from LLMs wrapped in tool-using harnesses rather than from stand-alone models. Much of the current discourse around them starts from a shrinking narrative. Foundation models improve, agents can browse repositories and invoke tools over longer horizons, and tasks such as scaffolding, routine test generation, straightforward bug fixing, and small integration work look more exposed than they did only a few years ago. Natural-language interfaces also make it easier for end users and domain experts to assemble working solutions without writing as much traditional code. The result is anxiety not only among students and junior developers, but also among experienced practitioners who wonder how much of their hard-won expertise will retain value. Empirical signals support both that anxiety and the need for caution about simplistic replacement stories: early-career workers in the most AI-exposed occupations have already seen measurable employment pressure \citep{stanford2025canaries}, but realistic software work still reveals important limitations in current tools \citep{metr2025osdev}. Those concerns are real, and dismissing them as temporary noise would be a mistake.
The argument below is that those concerns, taken seriously, point not to a shrinking field but to a broader engineered object: software engineering increasingly spans \emph{semi-executable} artifacts---prompts, workflows, policies, and organizational routines that are not compiled or executed in the classical sense, yet are read and acted on by models, agents, and people at runtime, and therefore shape system behavior as directly as code does.

The more important question is what follows if some of those concerns are correct. The central claim of this paper is that they do not point to a smaller role for software engineering. They point to a broader engineered object. As systems become more agentic, more of what organizations do is specified and coordinated through artifacts that are neither classical code nor fully informal human routines. Prompts, workflows, policies, evaluation harnesses, routing logic, escalation rules, and decision procedures increasingly shape system behavior in practice. These artifacts are not executable in the classical sense, but neither are they mere documentation. They occupy a middle ground in which behavior is partly specified, partly interpreted, and often consequential.

\begin{namedbox}{Thesis: The Expanding Scope}
AI-based agentic systems do not make software engineering smaller. They expand the engineering object outward along a spectrum of semi-executability, from code and tests toward prompts, workflows, controls, organizational operating logic, and eventually societal and institutional fit.
\end{namedbox}

This reframing matters because capability claims, adoption claims, and organizational-change claims are not the same thing. Capability progress can be rapid without immediate institutional change. Adoption can spread quickly in narrow settings while remaining unreliable in complex ones. Organizational change is slower still because it runs into trust, incentives, politics, compliance, procurement, and habits \citep{svensson2019unfulfilled}. Slowness, however, does not mean irrelevance. It means the transition is mediated by socio-technical engineering rather than decided by benchmarks alone.

Three supporting observations make the expansion argument concrete. They explain why imperfect, unevenly capable agentic systems can still reshape what organizations engineer, and why the expansion holds even where individual model performance remains weak.

\textbf{Utility, not perfection.} The bar for triggering organizational change is not parity with the best human engineer, but adequacy against the real allocation of scarce expert attention inside actual organizations. Systems that are useful enough and cheap enough can reshape what gets built or automated even when they fall well short of expert performance on any single dimension \citep{metr2026timehorizons,google2024aise}.

\textbf{Compounding availability beats isolated excellence.} The Utility point sets where the adequacy bar sits; the compounding point is about what happens once that bar is cleared. Thousands of low-friction uses that reshape everyday routines compound into organizational capability that sporadic access to a top expert cannot match, which is one reason real-world use already concentrates in software-related work even while model performance remains uneven \citep{anthropic2026economic,metr2025osdev}.

\textbf{Broader creation multiplies the need for engineering.} If natural-language interfaces let more domain experts, analysts, and operators create useful systems, the likely result is not less engineering. It is more variation in quality, more hidden dependencies, and more need for reusable validation, governance, and lifecycle discipline. This echoes a long-running lesson from end-user software engineering, where empowering non-specialist builders consistently raised rather than lowered the demand for engineering discipline around the artifacts they produced \citep{ko2011enduser,lieberman2006enduser}. Lower barriers to creation therefore enlarge the deployment surface that engineering must support, even when some classical programming bottlenecks weaken.

Together, these three observations explain why the engineered object expands even before any single agentic system reaches parity with expert human engineers: utility, availability, and broader creation each push work outward along the stack.

To see why this matters, consider a simple release-preparation workflow. A team still has code, tests, and build artifacts. But it may also rely on prompts that summarize regressions, orchestration logic that gathers evidence from issue trackers and telemetry, evaluation harnesses that compare generated summaries with ground truth, escalation rules that decide when a human must intervene, and an organizational routine that determines who signs off and under what conditions. The outcome depends on how well those artifacts line up with one another, not on code quality alone. The same pattern is already visible in developer-facing systems such as Claude Code and Codex, where prompts, repository instructions, tool permissions, and reusable skills partially encode how work should proceed. A paper about agentic software engineering therefore needs an analytic vocabulary for the full surface being engineered. The stack applies symmetrically: it reasons both about how AI expands the engineered object of software engineering, and about how software engineering in turn engineers the AI systems themselves.

The argument in this note is informed by published industrial work in an automotive setting with Volvo-affiliated collaborators \citep{wang2025spapi,khoee2024gonogo}, and by earlier work on autonomous testing agents and intent-driven agentic GUI testing \citep{feldt2023autonomous,yoon2024intent}. The aim is a compact diagnostic model that software engineering researchers and sophisticated practitioners can use to ask where a contribution, bottleneck, or transition primarily sits.

As a conceptual keynote companion, this paper makes five contributions:
\begin{itemize}
  \item a tighter definition of \emph{semi-executable artifact};
  \item a six-ring diagnostic reference model, \emph{The Semi-Executable Stack};
  \item a demonstration of diagnostic use across three worked cases;
  \item five named reframing moves that treat common objections as engineering targets; and
  \item a preserve-versus-purify heuristic for deciding which legacy software engineering processes, controls, and coordination routines should be kept and which should be simplified or redesigned.
\end{itemize}

The paper does \emph{not} propose a maturity model in the Capability Maturity Model Integration (CMMI) sense \citep{chrissis2011cmmi}, a complete process framework, or an evaluation rubric. Its aim is diagnostic and agenda-setting rather than procedural completeness.

\section{The Semi-Executable Stack}

\begin{figure}[H]
  \centering
  \includegraphics[width=\linewidth]{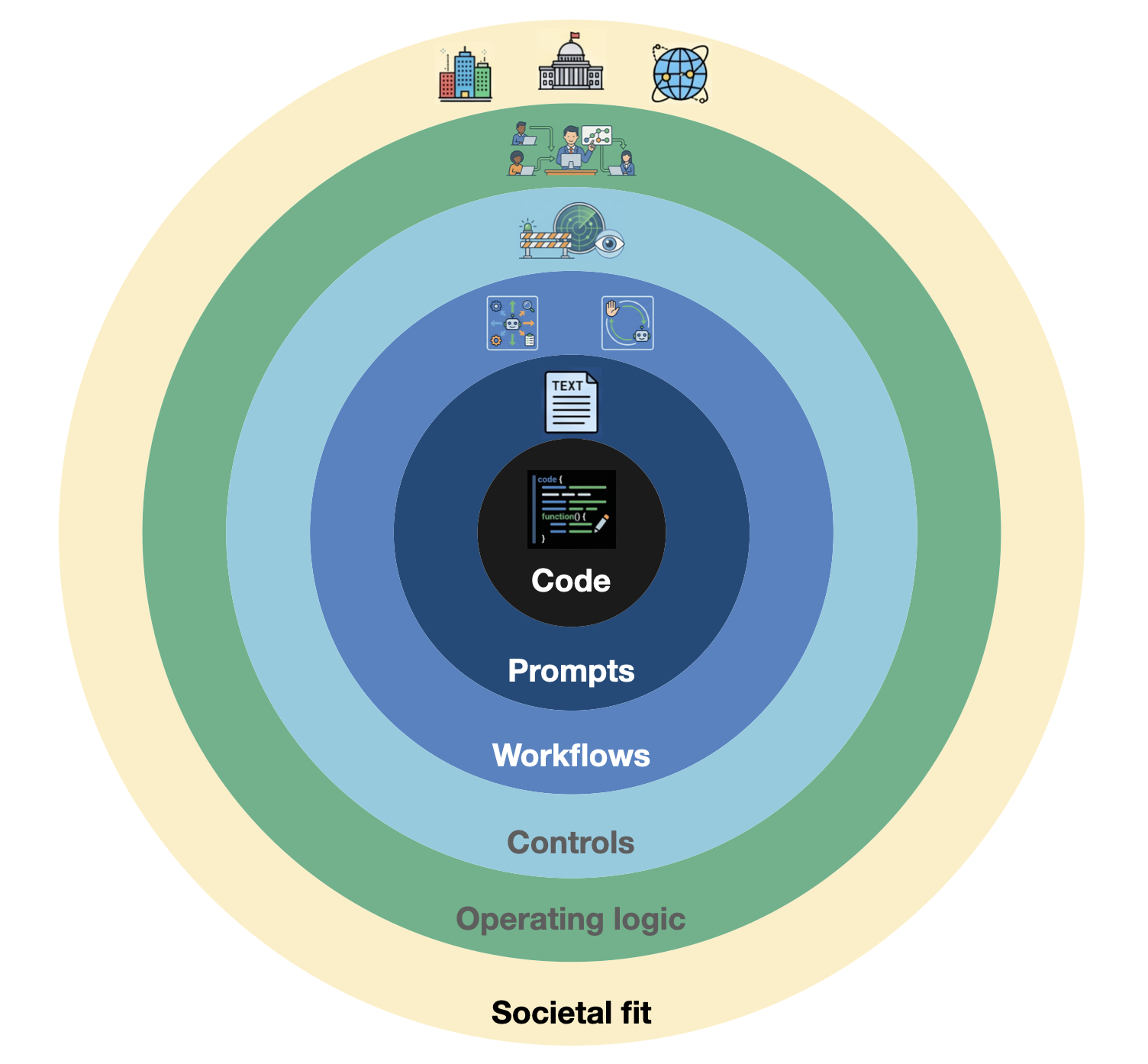}
  \caption{The Semi-Executable Stack. The rings represent a spectrum of engineering objects, not a sequential lifecycle. Inner rings are more fully specified and more directly executable; outer rings rely more heavily on interpreted, probabilistic, and organizational forms of execution. The short ring labels (Code, Prompts, Workflows, Controls, Operating logic, Societal fit) are representative examples used for brevity; the text uses fuller names (e.g.\ Executable artifacts, Instructional artifacts, Orchestrated execution) that better capture the scope of each ring.}
  \label{fig:stack}
\end{figure}

Figure~\ref{fig:stack} presents the paper's main reference model. The model uses \emph{rings} rather than \emph{levels} or \emph{layers} because the point is not a stepwise hierarchy or a stack of separable modules. The rings mark regions on a spectrum defined jointly by specification completeness and execution determinism, with blurry boundaries and strong adjacency effects. They should therefore not be read as phases, maturity levels, or runtime layers. Inner rings are closer to classical executable software. Outer rings involve more interpretation by humans, probabilistic models, or both.

\begin{namedbox}{Definition: Semi-Executable Artifact}
A semi-executable artifact is a software-related artifact that helps specify, coordinate, or constrain the behavior of a software-intensive system and its surrounding workflow, but whose enactment depends on interpretation by humans, probabilistic models, or a combination of the two, rather than on fully deterministic machine execution alone.
\end{namedbox}

Semi- does not mean non-executable. It means that the artifact helps drive behavior, but does not by itself determine that behavior in the fully specified, deterministic sense associated with classical programs. Its execution is instead approximate, context-sensitive, and probabilistic, and it may also involve human judgment or intervention in some of its steps.

Code is the limiting case of high executability. Organizational routines are the limiting case of low executability. Between them sit prompts, natural-language specifications, workflow rules, policy layers, and decision procedures. A purely informal norm with no stable representational form sits outside the low end of the spectrum; a fully verified executable lies close to the inner extreme. The point is not that all of these artifacts are equivalent. It is that modern software engineering increasingly has to engineer across this spectrum rather than only at the executable-code end of it.

\begin{namedbox}{Model: The Semi-Executable Stack}
The Semi-Executable Stack is a six-ring reference model for reasoning about the expanding object of software engineering. Its primary purpose is diagnostic: to show which ring a contribution or bottleneck primarily inhabits, and which adjacent rings it depends on.
\end{namedbox}

The six rings are as follows.

\begin{enumerate}[leftmargin=1.6em]
  \item \textbf{Executable artifacts.} Code, tests, schemas, and configurations. What is engineered here is behavior through direct machine execution. Typical examples include generated test code, repositories, and CI configurations.
  \item \textbf{Instructional artifacts.} Prompts, natural-language specifications, task descriptions, and exemplars. What is engineered here is behavior through guided interpretation. SPAPI-Tester's prompt templates and API-to-test transformations fit here \citep{wang2025spapi,prompting2025wild}.
  \item \textbf{Orchestrated execution.} Tool use, retrieval, agent workflows, multi-agent protocols, and human-agent loops. What is engineered here is behavior through structured interaction among components. Multi-step test-generation pipelines and release-support workflows are examples.
  \item \textbf{Control systems.} Guardrails, monitoring, evaluation harnesses, policy layers, escalation rules, and failure-response protocols. What is engineered here is acceptable operation, observability, and correction, in the sense emphasized by the National Institute of Standards and Technology (NIST) AI risk-management framing \citep{nist2023rmf,nist2024genai}. A weekly release gate that checks agent output quality and routes uncertain cases to human review sits here.
  \item \textbf{Operating logic.} Decision preparation, coordination routines, knowledge capture and reuse, delegation structures, and role exchange. What is engineered here is recurring organizational behavior: the organization's operating logic, that is, its organizational logic as enacted in practice \citep{dora2025state,cybernetic2025}. GoNoGo and \emph{The Cybernetic Teammate} are examples of this ring \citep{khoee2024gonogo,cybernetic2025}.
  \item \textbf{Societal and institutional fit.} Cross-organizational integration, institutional legitimacy, public-sector adoption, regulatory fit, and broader societal embedding. This outer ring constrains what viable engineering looks like even when it is not itself engineered as directly as rings 2--5. The EU AI Act is a current example of this outer pressure \citep{eu2024aiact}.
\end{enumerate}

Historically, software engineering has centered on ring 1 and partly on ring 2. What is changing is that rings 2--5 increasingly become first-class engineering objects, while ring 6 increasingly constrains what succeeds in practice. This is not a claim that every organization will immediately engineer all six rings well. It is a claim about where important bottlenecks and design choices increasingly reside. In general, the outer rings are more semi-executable because they embed more tacit knowledge, judgment, trust, politics, and context, and therefore depend more heavily on human interpretation even when parts of them become formalized.

The model is intended for diagnosis, not classification for its own sake. A contribution's \emph{primary ring} is where its main engineered artifact sits; in practice, the primary ring is identified by asking: \emph{if this ring's artifact were removed or poorly designed, would the system's value collapse?} A contribution is genuinely \emph{multi-ring} when its novelty is split across rings rather than merely depending on adjacent rings for deployment. ``Moving outward'' means that the bottleneck, evaluation criteria, and durable organizational effect increasingly sit farther from fully executable code.

\begin{table}[t]
  \centering
  \small
  \caption{Using the stack diagnostically.}
  \label{tab:diagnostic}
  \begin{tabularx}{\linewidth}{P{0.34\linewidth} Y}
    \toprule
    \textbf{Question} & \textbf{Diagnostic use} \\
    \midrule
    What is the main engineered artifact? & Assign a tentative primary ring. If the novelty is mainly in prompts, workflows, or routines rather than generated code, the contribution likely sits outside ring 1. \\
    Where does execution depend most on interpretation? & Distinguish executable, instructional, orchestration, and operating-logic contributions. Higher dependence on human or model interpretation usually marks movement outward. \\
    Where do the main evaluation criteria sit? & Reveals whether success is about local capability, control quality, organizational routine change, or institutional legitimacy. \\
    Which human unit of analysis carries the main change? & Makes the behavioral impact explicit rather than leaving it implicit. Following behavioral software engineering, ask whether the contribution mainly changes individual cognition and judgment, team interaction and coordination, or organizational routines and incentives \citep{lenberg2015bse}. \\
    Which adjacent rings are structurally necessary for deployment? & Marks multi-ring dependence without collapsing everything into one layer. Most useful systems depend on multiple rings even when one ring dominates the contribution. \\
    \bottomrule
  \end{tabularx}
\end{table}

Table~\ref{tab:diagnostic} deliberately adds a behavioral question to an otherwise artifact-centered model. This is where the stack connects to behavioral software engineering: the rings help identify what is being engineered, while BSE reminds us to ask which people and social units absorb the change. Many agentic systems do not merely alter artifacts. They also redistribute attention, judgment, coordination, accountability, trust, and perceived professional agency across individuals, teams, and organizations \citep{lenberg2015bse,lenberg2015humanfactors,fagerholm2022cognition,graziotin2022psychometrics}. In that sense, BSE is best read here as a cross-cutting lens rather than as an additional ring: the stack diagnoses where the engineering work sits, while BSE helps explain how that change is experienced and negotiated by the humans around it.

Ring 6 deserves one concrete anchor because it is easiest to overstate. The point is not that regulation or public legitimacy suddenly become software engineering in the same direct sense as code review or testing. The point is that outer-ring pressures now reshape what inner-ring designs are viable. The EU AI Act makes this concrete: obligations around risk management, logging, transparency, human oversight, and post-market monitoring for high-risk AI systems reach inward and change ring-4 control artifacts and ring-5 operating routines \citep{eu2024aiact}. In regulated domains such as automotive, analogous sectoral expectations around safety and accountability create the same directional pressure even when the exact mechanism differs.

This model differs from adjacent views in several ways. It is not a lifecycle model such as waterfall because its rings are not phases in time. It is not a software architecture stack because the rings are not runtime layers inside a deployed artifact. And it is not simply another generic socio-technical framing. The contribution here is the explicit focus on a spectrum of artifacts whose execution now mixes deterministic software, probabilistic models, and organizational interpretation. That spectrum matters because it changes where engineering leverage sits.

The language of \emph{semi-executable} names the destination concept. The language of \emph{expanding} names the dynamic. The historical trend has been outward. Around 2015, most software engineering work still centered on source code, tests, build pipelines, tickets, and relatively static requirements documents. Today, prompts live in repositories, evaluation harnesses gate releases, workflow logic coordinates human-agent loops, and policy layers increasingly shape deployment. Many artifacts once treated as documentation, process residue, or management overhead are being represented in forms that can be inspected, executed approximately, monitored, and revised with software-like discipline \citep{prompting2025wild,microsoft2025frontier}.

\section{Positioning Relative To Prior Work}

The Semi-Executable Stack is intended to sit alongside, and partly synthesize, several adjacent literatures rather than replace them. Unlike lifecycle models, architecture stacks, or socio-technical framings that treat organizational context as background, the stack's distinguishing move is to organize contributions along a spectrum of \emph{artifact executability}---from deterministic code to interpreted organizational routines---and to treat every point on that spectrum as an engineering object. This section therefore does not aim at exhaustive coverage. Its purpose is to position the paper more precisely. An earlier framing considered a narrower range of AI applications to software engineering, largely at the inner rings \citep{feldt2019waysai}; the present stack generalizes that earlier map now that the engineered object itself has broadened outward.

\smallskip\noindent\textbf{AI4SE and agentic-coding benchmarks.}\ Recent LLM4SE surveys and agent-for-SE surveys provide useful maps of the current technical landscape \citep{fan2023survey,hou2024slr,liu2024agentse}. Their center of gravity lies in rings 1--3: code generation, bug fixing, testing, repair, prompt engineering, agent design, and benchmark-driven capability progress. Benchmarks such as SWE-bench extend that frontier toward longer-horizon, repository-level issue resolution \citep{swebench2024}, making them useful indicators of progress at the executable and orchestration rings. The stack complements these contributions by making it easier to locate which problems sit at the control, operating-logic, and institutional rings.

\smallskip\noindent\textbf{Prompts, plans, and agentic artifacts.}\ Prompt-engineering and prompt-as-artifact work speaks directly to ring 2. \emph{Prompting in the Wild} is especially useful because it shows prompts evolving in repositories as maintained, though still weakly disciplined, engineering artifacts \citep{prompting2025wild}. Earlier work on conversational testing agents and intent-driven agentic GUI testing also sits in this inner-to-middle region of the stack: it demonstrates genuinely agentic software engineering without yet claiming outer-ring organizational redesign \citep{feldt2023autonomous,yoon2024intent}. Contemporary developer systems such as Claude Code and Codex also make this shift visible in practice: prompts, skills, tool permissions, and repository instructions increasingly encode reusable workflow structure rather than merely wrapping code generation. Read together, these examples reveal a small but useful trajectory: from early autonomous agents operating against software artifacts and interfaces toward later industrial systems in which prompts, plans, workflows, and control logic become first-class engineering objects. The stack builds on that insight but generalizes it. Prompts and agent plans matter not only because they are new artifact classes, but because they make visible that the engineered surface now extends beyond code.

\smallskip\noindent\textbf{MLOps, SE4ML, SE4AI, and evaluation.}\ MLOps and adjacent LLMOps framings are close neighbors for ring 4. Their main contribution is to treat deployment, monitoring, reproducibility, and operational control as first-class concerns rather than afterthoughts \citep{kreuzberger2023mlops}. The stack agrees with that move, but pushes the boundary further outward. MLOps usually centers the operation of models and pipelines. Recent work such as GateLens likewise shows how formal intermediate representations can improve traceability and reliability in LLM-based release analytics \citep{khoee2025gatelens}. Holistic evaluation approaches for language models add another ring-4 building block by treating multi-metric, scenario-conditioned evaluation as an engineered artifact rather than a one-off benchmark score \citep{liang2023helm}. The present argument is that the engineering surface now extends further still, to decision routines, coordination structures, and external legitimacy constraints.

The SE4ML literature established that ML systems accumulate hidden technical debt outside code, in data dependencies, feedback loops, and configuration \citep{sculley2015hiddendebt,giray2021seml}; once the engineered object expands to agentic operating logic, similar debt accumulates at rings 3--5 as well, in prompts, plans, evaluation harnesses, and release-decision routines. A parallel SE4AI perspective treats the engineering of agentic and LLM-based systems themselves---their prompts, tools, guardrails, and evaluation harnesses---as an SE concern in its own right \citep{martinezfernandez2021se4ai}, and the stack applies symmetrically to those systems.

\smallskip\noindent\textbf{Socio-technical and behavioral SE.}\ The paper sits in a socio-technical systems tradition that already argued software-intensive systems cannot be understood purely as technical artifacts \citep{baxter2011sts}. It also sits adjacent to behavioral software engineering, which emphasizes cognitive, behavioral, and social aspects of software engineering at the level of individuals, groups, and organizations \citep{lenberg2015bse}. This paper shares that premise: software engineering is shaped not only by technical structures, but also by human judgment, coordination, trust, and organizational context. What it adds is a more explicit account of the artifacts through which those socio-technical dynamics are increasingly expressed and enacted. Prompts, policies, orchestrations, and routines matter here not just as surrounding context, but as semi-executable artifacts that increasingly carry coordination, constrain behavior, distribute judgment, and shape how people and organizations interact with the system. Recent work on the socio-emotional and functional dimensions of human-AI collaboration in software engineering shows that these dynamics also have affective and behavioral facets that need to be engineered rather than assumed away \citep{muralirani2026socioemotional}.

\smallskip\noindent\textbf{Governance and institutional fit.}\ Work on human-agent teams, organizational amplification, and AI governance helps fill rings 5 and 6. DORA frames AI as an amplifier of existing organizational strengths and weaknesses \citep{dora2025state}. \emph{The Cybernetic Teammate} shows that AI can reshape how expertise is combined and how teams work together \citep{cybernetic2025}. Complementary empirical work on AI transformations of software organizations examines how such change actually unfolds in practice, from structural reconfiguration \citep{feldt2025orgchange} to cross-functional task forces designed to move AI capability into the organization without flattening local engineering judgment \citep{feldt2025xfaits}. NIST's AI RMF and the EU AI Act show that governance and institutional fit increasingly shape viable system design \citep{nist2023rmf,eu2024aiact}. The stack is intended as a compact reference model that positions these literatures relative to one another rather than leaving them conceptually separate.

Taken together, these literatures cover much of the ground the stack spans, but each addresses a subset of rings. The stack's contribution is not new empirical ground. It is the diagnostic frame that lets researchers and practitioners locate where a contribution sits, what it depends on, and what it leaves unaddressed.

\section{Three Worked Cases}

The stack is most useful when it separates contributions that would otherwise collapse into one broad category of ``AI for software engineering.'' Table~\ref{tab:cases} works through three such cases: two from a shared automotive setting, and one external contrast case that tests whether the framing travels. The cases are illustrative, not exhaustive. They show that the stack distinguishes what is actually being engineered.

\begin{table}[t]
  \centering
  \small
  \caption{Contrasting cases on the Semi-Executable Stack.}
  \label{tab:cases}
  \begin{tabularx}{\linewidth}{P{0.19\linewidth} P{0.11\linewidth} P{0.23\linewidth} Y P{0.16\linewidth}}
    \toprule
    \textbf{Case} & \textbf{Primary ring(s)} & \textbf{Engineering object} & \textbf{What it contributes} & \textbf{What remains unresolved} \\
    \midrule
    SPAPI-Tester \citep{wang2025spapi} & 2--3 & Prompted test-process workflow from API specifications to tests and code & Automates a substantial testing pipeline and contributes a recipe for preserving process structure while replacing decomposed manual steps & Trust, release integration, and broader organizational uptake remain largely outside the core contribution \\
    GoNoGo \citep{khoee2024gonogo} & 5, supported by 3--4 & Semi-executable release-decision support routine & Re-engineers decision preparation rather than only a coding-adjacent activity; encodes coordination, information flow, and bounded analysis logic & Broader governance, legitimacy, and long-term organizational embedding remain open \\
    The Cybernetic Teammate \citep{cybernetic2025} & 5 & AI-mediated combination of expertise and task coordination & Shows AI can change how expertise is combined and how work is coordinated even outside software generation & Domain transfer to software settings and institutional embedding still need study \\
    \bottomrule
  \end{tabularx}
\end{table}

\textbf{Case A: SPAPI-Tester.} Wang et al.\ \citep{wang2025spapi} automate a substantial automotive testing workflow, moving from API specifications to test cases and then to test code. The sequence matters: the engineered object is not a new foundation model but a semi-executable pipeline of prompts, transformation steps, workflow logic, and the generated test code they produce. This is primarily a capability claim with some local adoption significance, and it shows that ring-2 and ring-3 artifacts are engineered objects in their own right, not disposable glue around generated code.

Unlike many inner-ring automation papers, SPAPI-Tester is explicit about what makes the automation feasible: preserve the existing process structure, decompose it into verifiable steps, and replace only the manual steps with LLM-supported automation. Prompts and workflows become the engineering object, but no grand organizational redesign is required. The empirical results are concrete: experiments on more than 100 APIs, and a real-world run on 193 newly developed APIs that surfaced 23 failures---22 confirmed implementation bugs and one documentation-parsing problem. The contribution's limits sit at a different ring from its technical strength. The pipeline improves a coding-adjacent activity but leaves ring-4 and ring-5 questions open: who trusts its outputs, how those outputs enter release decisions, how exceptions are handled, and how surrounding roles change. The stack makes that limit visible without diminishing the contribution, and it shows why the next step is better control and integration, not simply ``better generation.''

\textbf{Case B: GoNoGo.} Khoee et al.\ \citep{khoee2024gonogo} use an LLM-based multi-agent system to support automotive software release decisions. A go/no-go routine is not merely a technical check. It is an organizational process for gathering status from multiple sources, synthesizing evidence, surfacing exceptions, and preparing a sign-off decision under time pressure. The system therefore encodes part of release-decision preparation---specific, structured analysis of tabular test data---rather than a code-generation task. The contribution sits at ring 5, supported by rings 3 and 4.

Judging GoNoGo only by the quality of the analysis scripts it generates would miss the contribution. Those scripts exist, but they are a byproduct. What is re-engineered is semi-executable operating logic: which signals matter, how they are assembled, how user queries become bounded analysis steps, and where human judgment is invoked. The paper's own evaluation matches that reading: perfect performance on simpler tasks up to Level 2 in a 3-shot setting, with human intervention recommended for more complex Level 3 and 4 tasks. That balance strengthens rather than weakens the example. The system is already deployed internally, reportedly saves pilot users about two hours per decision occasion, and offloads routine analytical work while leaving higher-level interpretation and final judgment with release managers. The stack explains why the evaluation criteria must differ from those of a ring-2/3 contribution. This is already an organizational-change claim, not only a capability or adoption one.

\textbf{Case C: The Cybernetic Teammate.} The two automotive cases share teams and industrial context, which holds much of the domain constant while shifting the main engineering object outward. The third case serves a different purpose: testing whether the stack travels beyond that local setting. It also illustrates how the stack should be used. The point is not to assign a contribution to a single ring, but to locate where the main novelty and the main unresolved bottlenecks sit. Two misuses are equally tempting: flattening the cases back into one ``AI for SE'' category, or ignoring the multi-ring dependencies that make each viable.

A field experiment with 776 professionals at Procter \& Gamble found that individuals using AI matched the performance of teams without AI and showed less functional siloing \citep{cybernetic2025}. This is not a software-generation result. It is evidence that AI can reshape how expertise combines and how decisions are prepared---the signature of ring-5 change. Whether the same effect transfers to software settings is an open question; stating that honestly matters. Even so, the case shows two things: that code need not even be part of the contribution, and that the stack is not merely a way of relabeling a local automotive portfolio.

Read together, the three cases trace a pattern. As the primary engineering object moves outward---from code (SPAPI-Tester) through decision preparation (GoNoGo) to expertise coordination (The Cybernetic Teammate)---the evaluation criteria, the skills required, and the unresolved bottlenecks all shift with it. The stack does not rank these contributions. It makes the shift visible and helps identify what each leaves open.

\section{Counters Become Engineering Problems}

The most common objections to the thesis identify real problems and cannot simply be waved away. Reliability failures are real. Maintenance debt is real. Organizational inertia is real. Power and politics are real. High-level judgment does appear harder to automate than lower-level coding work. But none of these objections imply that software engineering matters less. They show where the work moves and what kind of engineering becomes more central.

\begin{table}[t]
  \centering
  \small
  \caption{Common counters, where they bite, and the engineering move they imply.}
  \label{tab:counters}
  \begin{tabularx}{\linewidth}{P{0.29\linewidth} P{0.16\linewidth} Y}
    \toprule
    \textbf{Counter} & \textbf{Bites at rings} & \textbf{Engineering move} \\
    \midrule
    \textbf{Reliability failures}: agents hallucinate, miss context, and fail unpredictably & 1--3 & Reliability Becomes Engineering \\
    \textbf{Maintenance debt}: lower-cost, faster generation yields fragile dependencies and shallow understanding & 2--4 & More Creation Requires More Maintenance Discipline \\
    \textbf{Organizational inertia}: incentives, procurement, compliance, and habits slow uptake & 4--5 & Adoption Is Engineering \\
    \textbf{Power and politics}: redistribution of expertise and control is resisted & 4--6 & Governance Is Engineering \\
    \textbf{Irreducible judgment}: architecture, sensemaking, and taste resist automation longer & 5--6, but felt throughout & Centre of Gravity Moves Outward \\
    \bottomrule
  \end{tabularx}
\end{table}

\begin{namedbox}{Reliability Becomes Engineering}
Once agentic systems are useful enough to deploy, verification, monitoring, evaluation harnesses, and oversight stop being peripheral wrappers. They become core engineering artifacts in their own right.
\end{namedbox}

The right response to unreliable agents is not to declare the transition illusory. It is to recognize that control systems move to the center of the engineering problem. A common failure mode is silent misalignment: a model summary looks plausible, a retrieval component changes, or a tool call drifts, and the problem is discovered only after later decisions have already absorbed it. The answer is not a better slogan about trust. It is a set of concrete artifacts: evaluation harnesses, coverage checks, monitoring dashboards, escalation thresholds, and human-oversight loops. Architectures such as GateLens show one practical response by narrowing the reasoning-to-code gap through intermediate representations that are easier to inspect and verify \citep{khoee2025gatelens}.

\begin{namedbox}{More Creation Requires More Maintenance Discipline}
Every increase in the capacity to create software-like artifacts increases the need for architecture, traceability, and lifecycle discipline. Faster creation has historically produced more maintenance work, not less.
\end{namedbox}

This point applies well beyond code. Prompt changes, policy changes, workflow changes, and coordination changes all accumulate debt when they are treated as disposable rather than engineered \citep{prompting2025wild}. The problem is not only speed. It is that these artifacts become cheap enough to produce in larger numbers and with less deliberation. A concrete failure mode is prompt drift: a prompt is updated to improve one case, downstream behavior shifts, and no one can later reconstruct why the system changed. The engineering response is lifecycle discipline pushed outward: version prompts and workflow logic, track evaluation suites alongside code, and preserve traceability from intent to behavior. Similar brittleness and robustness trade-offs are well known in test automation work, where small interface shifts can break scripts unless locator and matching strategies are engineered explicitly \citep{nass2023similarity}. Recent taxonomy work on testing LLM-based software sharpens the point by showing that correctness often has to be assessed across repeated runs and varying configurations rather than as a binary property of a single execution \citep{dobslaw2026taxonomy}. The maintenance burden therefore expands with the stack.

\begin{namedbox}{Adoption Is Engineering}
If the bottleneck is organizational rather than technical, then transition design, legitimacy, coordination, and role adaptation become engineering concerns rather than optional change-management extras.
\end{namedbox}

This is one reason DORA's amplifier framing matters. AI does not inject capability into a neutral environment. It interacts with existing process quality, incentives, and system clarity \citep{dora2025state}. A common failure mode is easy to recognize: a technically promising tool is dropped into a workflow with unclear ownership, weak exception handling, or no agreed sign-off path, and it produces confusion rather than leverage. The engineering response is a set of transition artifacts: operational playbooks, role definitions, review checkpoints, and coordination routines that let new capability enter real work \citep{alegroth2013transitioning}. In practice, the bottleneck often lies in how capabilities enter the organization, not in whether a benchmark number rises \citep{svensson2019unfulfilled}. This is where adoption claims turn into organizational-change claims.

\begin{namedbox}{Governance Is Engineering}
Power, accountability, legitimacy, and institutional fit are part of the engineering context for semi-executable systems. Treating them as purely external constraints is itself a design mistake.
\end{namedbox}

Once semi-executable artifacts start shaping decisions, delegation, and coordination, they also inherit disputes about ownership, auditability, acceptable risk, and control. These are not abstract social add-ons. They shape what the system can be trusted to do and which rings can be expanded in practice. The EU AI Act makes this concrete because obligations around logging, oversight, risk management, and post-market monitoring can force changes in system architecture, review paths, and operational routines \citep{eu2024aiact}. Empirical work on decision-making in responsible SE for AI shows that practitioners already treat these obligations as engineering decisions rather than as compliance overhead, even when organizational support remains thin \citep{muralirani2025decisionmaking}. The response is therefore not only technical hardening. It is also the design of accountable artifacts: audit trails, escalation rules, model cards, approval boundaries, and evidence bundles that support legitimate use.

\begin{namedbox}{Centre of Gravity Moves Outward}
Even as AI systems improve, judgment, architecture, and sensemaking remain harder to formalize and automate than many lower-ring tasks. That is exactly why those activities become more central as lower-ring tasks become cheaper and more automated.
\end{namedbox}

This is the most important response to the objection that ``AI cannot do the highest-level work.'' The point is not that the outer rings remain untouched. It is that the scarcest human work increasingly sits in specification, architecture, judgment, evaluation, and organizational fit precisely because some lower-ring production becomes cheaper. A practical failure mode is false substitution: organizations automate local coding tasks but neglect the harder work of defining intent, validating behavior, and redesigning routines. The engineering response is to shift attention toward outer-ring artifacts rather than treating them as non-engineering residue. In that sense, the center of gravity of software engineering moves outward even if some lower-level work becomes easier or more automated.

Each objection identifies a ring where more explicit design, validation, or organizational work is required. The reframing does have limits. In domains where failures are catastrophic and irreversible, where the required ring-4 controls cannot be built with current techniques, or where organizational legitimacy is actively contested, the honest conclusion may be that some agentic deployments should not happen yet. The stack does not guarantee that every semi-executable system is worth engineering. It helps identify when engineering across the full spectrum is feasible and when it is not.

\section{Preserve and Purify}

The expanding-scope thesis is not a license to throw out everything inherited from software engineering. Some long-standing practices capture durable principles. Others mainly reflect older tools, lower-bandwidth coordination, and a world in which most software was still produced manually.

\begin{table}[t]
  \centering
  \small
  \caption{A preserve-versus-purify heuristic for agentic software engineering.}
  \label{tab:preserve-purify}
  \begin{tabularx}{\linewidth}{Y Y}
    \toprule
    \textbf{Preserve} & \textbf{Purify} \\
    \midrule
    Explicit reasoning about requirements, constraints, and assumptions & Practices optimized mainly for manual code production \\
    Modularity and interface discipline & Process artifacts created primarily for low-bandwidth human coordination \\
    Validation and verification & Code-centric notions of quality that ignore prompts, policies, and workflows \\
    Traceability between intent, implementation, and behavior & Strict phase boundaries inside interactive human-agent loops \\
    Lifecycle thinking and maintenance awareness & Narrow views of who counts as a system builder \\
    Socio-technical realism about teams, communication, and failure & Techno-focus and techno-optimism that ignore organizational fit \\
    \bottomrule
  \end{tabularx}
\end{table}

\begin{namedbox}{Heuristic: Preserve vs. Purify}
Preserve the principles that remain necessary when the engineering object expands. Purify away the accidental complexity that was optimized for older tools, older coordination limits, and a narrower notion of what counted as software.
\end{namedbox}

In Brooks's classic sense, the goal is not to wish away essential complexity. It is to strip away accidental complexity that came from older tools, interfaces, and coordination constraints \citep{brooks1987silverbullet}.

The first theme is \textbf{specification and interfaces}. Explicit specifications, documented assumptions, modularity, and interfaces still matter because the expanding stack does not remove ambiguity; it moves it around. Specifications now cover intent for semi-executable behavior as well as code, and interfaces now include prompts, policies, and workflow boundaries alongside APIs and modules.

The second theme is \textbf{verification, traceability, and lifecycle thinking}. As generated and semi-generated artifacts proliferate, traceability now means linking prompts, retrieved context, policy settings, and observed outcomes, not only connecting requirements to code. For LLM-based software, variability-aware and aggregated oracles become necessary because repeated executions may differ even when the intended behavior is stable \citep{dobslaw2026taxonomy}.

The third theme is \textbf{socio-technical realism and a broader builder boundary}. Prompt authors, workflow designers, evaluators, operators, and domain experts increasingly co-create system behavior alongside software engineers. Some inherited process residue should be purified away because it compensated for historically low-bandwidth human coordination rather than expressing a durable principle: strict phase boundaries are the clearest example.

Two diagnostic questions make the heuristic more concrete. First: \emph{does this practice encode a durable principle (e.g.\ traceability, separation of concerns, accountability), or does it mainly compensate for a constraint that no longer binds (e.g.\ slow builds, low-bandwidth communication, manual code production)?} Second: \emph{when the engineered object expands outward to prompts, workflows, or operating routines, does the practice extend naturally to those artifacts, or does applying it there create friction without commensurate benefit?} The first question helps identify what to preserve. The second helps identify what to purify or redesign.

\section{Discussion}

The discussion looks at three angles: the research agenda the stack suggests, implications for practitioners, and current limitations.

\subsection{A research-agenda map}

The stack intersects both AI4SE research---using AI to support software engineering tasks---and SE4AI research---engineering AI-based systems as the primary concern \citep{martinezfernandez2021se4ai}. Most AI4SE work concentrates on rings 1--3: code generation, bug fixing, testing, repair, and benchmark-driven capability progress \citep{fan2023survey,hou2024slr,liu2024agentse,swebench2024}. SE4AI work addresses more of the outer rings, but the emphasis has largely been on evaluation, MLOps, and pipeline operation at ring 4 \citep{kreuzberger2023mlops,liang2023helm}. If the stack's argument holds, SE4AI for rings 5 and 6---decision routines, operating logic, governance, and institutional fit---represents the most underserved frontier. DORA's account of organizations as amplifiers \citep{dora2025state}, Microsoft's framing of human-agent teams \citep{microsoft2025frontier}, and results such as \emph{The Cybernetic Teammate} \citep{cybernetic2025} all point toward that gap. Table~\ref{tab:agenda} maps it by ring. This forces a sharper question for any given contribution: is it advancing a stable engineering frontier, or refining a legacy bottleneck that agentic systems are already beginning to absorb?

\begin{table}[t]
  \centering
  \small
  \caption{A compact research-agenda map suggested by the stack.}
  \label{tab:agenda}
  \begin{tabularx}{\linewidth}{P{0.06\linewidth} P{0.26\linewidth} Y P{0.22\linewidth}}
    \toprule
    \textbf{Ring focus} & \textbf{Current concentration} & \textbf{Under-explored questions} & \textbf{Likely methods} \\
    \midrule
    1--3 & Code generation, repair, testing, issue resolution, agent design & Which inner-ring gains actually transfer into durable workflow change? & Benchmarks, repository studies, field deployments \\
    4 & Evaluation, monitoring, guardrails, pipeline operation & What are the right abstractions for validating human-agent systems over time? & Longitudinal experiments, control/evaluation studies, incident analysis \\
    5 & Decision routines, coordination, role redesign & How do organizations move from local pilots to changed operating logic, and why do most fail? & Field studies, intervention studies, qualitative studies, experience sampling \\
    6 & Governance, legitimacy, regulatory fit, cross-organizational embedding & How do institutional constraints reshape viable technical and organizational designs? & Comparative case studies, policy analysis, socio-technical systems research \\
    \bottomrule
  \end{tabularx}
\end{table}

Beyond the artifact-level agenda captured in Table~\ref{tab:agenda}, a stronger human and organizational perspective is also needed. The question is not only which artifacts change, but how cognition, trust, identity, incentives, power, and coordination shift as those artifacts change. Prompting at ring 2 shifts human work from direct writing to specifying, steering, and validating AI-generated artifacts. Evaluation infrastructure at ring 4 shifts some responsibility for failure to the people who define how systems are assessed. Decision support at ring 5 redistributes whose judgment carries weight. Each is as much a question about human behavior and organizational dynamics as it is a technical one.

\subsection{Implications for practitioners}

The practical implication is not that every team should rush to automate everything. It is that engineering discipline matters even more as the object of engineering expands. The scarce skill shifts from building faster to deciding what is worth building or changing, which ring is actually being changed, how that change will be validated, how it will be governed, and how it will be maintained over time. Teams that treat AI only as a ring-1/2 efficiency tool may still gain local productivity, but they are also more likely to miss the deeper question of organizational redesign. Many AI initiatives fail in practice not because the underlying models are too weak, but because the organization misidentifies the ring it is really trying to change and therefore neglects the dependencies that matter most. A simple diagnostic can help: name the ring whose failure would kill the project, name the two adjacent rings it depends on, and compare those to the ring currently receiving most of the engineering effort. When the critical ring and the effort ring diverge, projects tend to stall.

For leaders, the implication is broader. AI is not best understood as a portfolio of local efficiency tools. The harder question is whether the organization’s operating logic is fit for a world in which more knowledge work is partially encoded, executed, evaluated, and improved through semi-executable artifacts. That is why waiting for certainty may itself be a losing strategy: while an organization waits for the tools to become unambiguously mature, competitors and internal bottlenecks may already be adapting around them.

In practice, this broader shift will usually be built through a sequence of narrower but credible wins at rings 1--3. Those wins accumulate the evidence and organizational patience needed for redesign at higher rings. Without leadership clarity about why those broader changes matter, however, the same wins are easily read as isolated productivity hacks rather than as stepping stones toward a different operating logic.

Semi-executability also changes the speed of learning. Fast feedback used to live mainly at ring 1; as prompts, workflows, and controls become semi-executable, organizations can iterate, compare, and decide across a broader engineering surface. The practical implication is to design higher-ring routines so that such learning happens by design rather than by accident.

This broader surface is also increasingly multidisciplinary: instructional artifacts come from domain engineers, controls from safety or compliance functions, and workflow logic from operational or managerial knowledge. The durable lesson for software engineers is to work more seriously with these neighboring disciplines, with both greater ambition and greater humility, rather than to conclude that outer-ring ownership displaces inner-ring discipline.

\subsection{Limitations}

The stack is a reference model rather than an empirical result, intended to support diagnosis and comparison rather than validation as a single experimental claim. The three worked cases are illustrative: two share an automotive industrial context, and the external contrast case demonstrates portability rather than full generalization. The evidence base for the outer rings is thinner than for the inner ones, so claims there are more directional than conclusive. Finally, ring assignment is often ambiguous---real systems span adjacent rings and reasonable analysts may disagree on which ring carries a contribution's primary novelty---and the stack is intended to support that discussion, not to adjudicate it.

\section{Closing}

The Semi-Executable Stack is offered as a compact diagnostic: not a prediction of a single future, but a map of where the engineering work is actually happening, where the bottlenecks sit, and why the hard work increasingly lies beyond code alone. The core argument is that AI-based agentic systems expand the engineering object rather than shrink the field. For researchers, the most underserved frontier is SE4AI for rings 5 and 6, where decision routines, operating logic, and institutional fit still lack the engineering methods the inner rings have accumulated. For practitioners, the implication is that prompts, workflows, evaluation harnesses, and decision routines deserve the same engineering discipline long applied to code, architecture, and tests.

\section*{Acknowledgements}

Robert Feldt is supported by the Chalmers Foundation's Academic Excellence Program (CAEP).

\bibliographystyle{plainnat}
\bibliography{references}

\appendix
\section{Keynote Title and Abstract}

\noindent\textbf{Title.} \emph{Agentic Software Engineering Will Eat the World: AI-Based Systems as the New Operating System of Society}

\medskip
\noindent\textbf{Abstract.}
AI-based and LLM-based systems are often cast as a threat to software engineering: foundation models improve, coding agents act through richer tool-using harnesses, some entry-level software work looks more exposed, and natural-language interfaces seem to let end users build useful systems without going through traditional software teams. This talk argues for a different conclusion. The deeper shift is not mainly about tools or tasks, but about organizational operating logic: how decisions are made, how knowledge moves, how work is coordinated, and how value is created. As these systems become more agentic, they will not merely support that shift; they will increasingly help analyze, adapt, and redesign it.

This change does not diminish the importance of software engineering. It broadens it. More of society will come to depend on semi-executable systems built from natural language, code, tools, policies, and workflows, while agentic environments increasingly generate, coordinate, and evolve many of those components. Concerns about reliability, maintainability, organizational inertia, and power are real. But these are better seen as constraints on the route than as arguments against the direction itself. Even imperfect AI can be transformative when deployed continuously at scale, and the central activities of software engineering are more likely to be redefined than replaced. For researchers, this creates a real risk that some AI4SE work targets practices already being folded into agentic workflows. For practitioners, it means that engineering discipline becomes more important, not less, as solution creation extends beyond traditional software developers.

\medskip
\noindent\textbf{Slide link.}
https://doi.org/10.5281/zenodo.19611576

\end{document}